Anthony C. Crawford        Saravan K. Chandrasekaran    Fermilab TD/SRF    acc52@fnal.gov     saravan @fnal.gov  25Jul16

# Demagnetization of a Complete Superconducting Radiofrequency Cryomodule: Theory and Practice

A significant advance in magnetic field management in a fully assembled superconducting radiofrequency cryomodule has been achieved and is reported here. Demagnetization of the entire cryomodule after assembly is a crucial step toward the goal of field less than $5\times10^{-7}$ Tesla. An explanation of the basic physics of demagnetization as well as experimental results are presented.

## Background

The recent advent of extremely low surface resistance, nitrogen doped niobium superconducting radiofrequency (SRF) cavities [1] has led to a redoubling of effort to prevent magnetic flux from being trapped in the niobium during transition from the normal to the superconducting state. Trapped magnetic flux in the thin layer that supports SRF current results in increased Ohmic losses that are the largest single contribution to residual resistance [2]. For continuous wave (CW) operation the added resistance causes a significant increase in thermal power that must be removed, at high cost, by a liquid helium refrigerator, typically operating at 2K. It is therefore possible for an accelerator project to realize considerable cost savings in equipment and operating funds by minimizing trapped magnetic flux.

## Introduction

Residual resistance due to trapped magnetic flux is larger for nitrogen doped SRF cavities than for pure niobium cavities, thereby increasing the importance of magnetic field management. Flux trapping in niobium is a complex phenomenon and a function of many variables. The most prominent sources for trapped flux and their appropriate means of mitigation are diagrammed in Figure 1. Information on the sources are to be found in the reference list: ambient field [3], magnetization [4], and thermocurrent [5], as well as for flux expulsion [6]. The term "thermocurrent" refers to electrical current caused by Seebeck effect voltage in the niobium cavity - titanium helium vessel electrothermal circuit, i.e., the thermocouple effect when temperature differences are present.

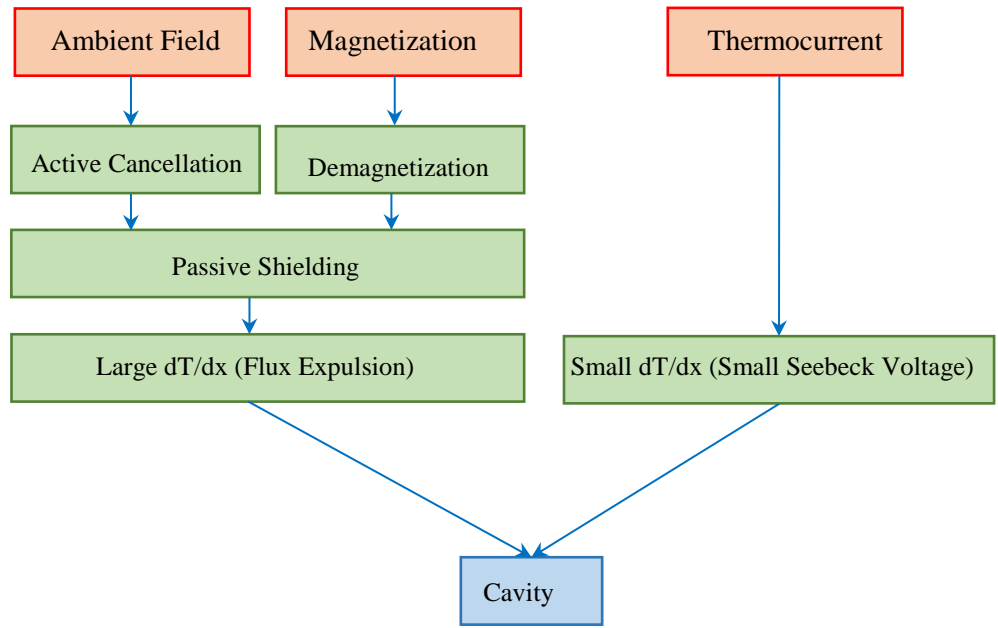

Figure 1    Sources of trapped flux and their means of mitigation. Sources are in the top row and mitigation is placed between the source and the cavity.

One particular difficulty, evident in Figure 1, is that achieving a small thermal spatial gradient (dT/dx) simultaneously with a large thermal spatial gradient may be problematic, although, in principle, it is possible to have a small gradient in the direction that causes thermocurrent and a large gradient in the orthogonal dimension. Efficient flux expulsion requires conditions that differ from those conducive to effective suppression of thermal current. Optimal arrangements for flux expulsion and thermocurrent management are being actively studied. Enough is known at the present time to allow the statement: If the thermocurrent electrical circuit can be broken by means of a ceramic insulator, thereby preventing any magnetic flux from thermal current, and if the niobium from which the cavity is made has received appropriate heat treatment to maximize flux expulsion, then a reasonably small value for dT/dx can be found that will result in essentially zero trapped flux in the cavity [7]. Until the time arrives when these techniques are mature, and are proved to be reliable, it safest to minimize the field at the cavity from ambient sources and magnetized components. Effective means of treating ambient sources can be found in [3]. The remainder of this report is devoted to the subject of demagnetization.

*Demagnetization*

In SI units, the relationship between applied magnetic field (**H**), magnetic field (**B**), and magnetic moment per unit volume (**M**) is:

$$(1) \qquad \mathbf{B} = \mu_0(\mathbf{H} + \mathbf{M})$$

This allows for clarification of the expression "Demagnetization". A material in a demagnetized state should have **B**=0 when **H**=0. According to Equation 1, this means that **M**=0 for a demagnetized material. In order for total magnetization to be the zero, the vector sum of all individual magnetization vectors from each magnetic domain in the material sample must be zero. While achievable under certain conditions, it will be shown that this state is not optimal for the purpose of magnetic shielding, including magnetic shielding for SRF cryomodules. In this report, only an object that is demagnetized according to the definition presented above, and that has zero remanence (**M**=0), will be referred to as demagnetized. An object that has undergone a demagnetization procedure, but that has remanent magnetization (**M**≠0), will be referred to with the use of Italicized letters: *demagnetized*.

One way of *demagnetizing* is to slowly diminish the magnitude of the AC current in the circuit shown in Figure 2. The object being *demagnetized* is a cylinder inside a solenoidal coil. For our case of special interest, the object would be a complete SRF cryomodule, approximately 11.4 meters in length and one meter in diameter. The geometry of our SRF cryomodule is that of ferromagnetic cylinders within ferromagnetic cylinders, all with their axes pointing in the same direction. So, the basic problem is that of a thin walled cylinder immersed in an axial magnetic field.

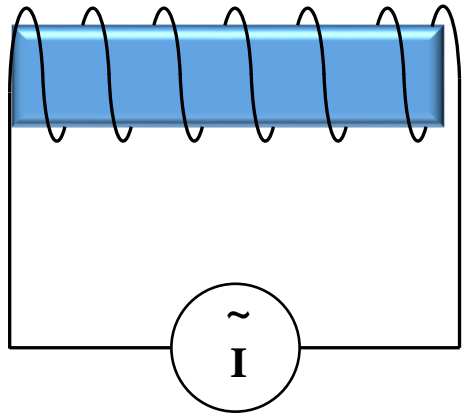

Figure 2    Concept for *demagnetization*

In practice, when we *demagnetize* a cryomodule, it is immersed in a DC ambient field due to the local Earth magnetic field, with its associated perturbations and irregularities. Thus the total applied field **H** has a decreasing AC component from a power source and a constant DC component from the environment, $\mathbf{H}_{DC}$. The effective applied field looks like Figure 3.

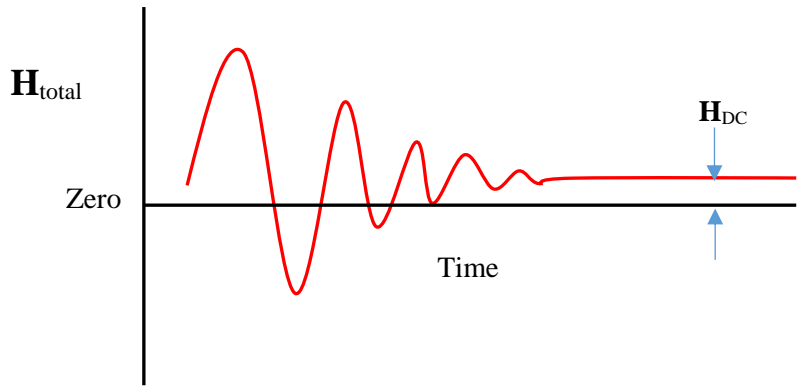

Figure 3    *Demagnetization* in an ambient field

If $H_{DC}$ were zero, then the cryomodule would, in principle, be truly demagnetized according to the definition presented previously.

Another way of *demagnetizing* is to heat the material to a state above the Curie temperature.  This, in addition to grain growth, is one of the goals of "magnetic annealing" performed by magnetic shield vendors.  If $H_{DC} = 0$ during the cooldown of the material, it will be demagnetized when it reaches room temperature.  If $H_{DC} \neq 0$ then the material will acquire overall magnetization as it cools, provided that the cooling is not faster than the time required for magnetic domain formation and ordering.  The reason that magnetization happens is that a partial ordering of magnetization vectors with the applied field minimizes the free energy of the magneto-thermodynamic system.  It is likely that commercially available, heat treated Cryoperm and Permalloy80 both have some amount of remanent magnetization following furnace treatment.  We will see that induced remanence due to $H_{DC}$ has a significant effect on apparent magnetic shielding efficiency.

**Modeling After Effects of *Demagnetization***

What happens after a thin walled, open ended cylinder is *demagnetized*?  A working model should include both ambient environmental magnetic field and remanent magnetization within the shield material.  A means of including remanence has been missing from previous cryomodule studies and will be developed here.

The approximate geometry for an ILC cryogenic magnetic shield, without end caps, is used here.  Applied field will be in the axial direction of the cylinder, as this is the most critical case for attenuation of field for SRF cavities.

Assume that $\mu_0 H_{DC}$ in Figure 1 is 0.0050 Tesla ( 500 milliGauss) in the axial direction of the cylinder, a value that is about the largest expected for a horizontal component of the Earth field.  The field is uniform when the magnetic material is not present.  What is the magnetic field distribution near the ferromagnetic material when it is immersed in the previously uniform field?  Results are shown in Figure 4 for the following values: cylinder inner radius = 0.127 m, wall thickness = 0.001 m, relative permeability = $\mu_r$ = 100,000.  The assumed value for $\mu_r$ is not unreasonably high for an undamaged cylinder made from annealed Cryoperm or Permalloy80.  There is symmetry under the operation of rotation about the Z axis in Figure 4.  For this simple model, the material permeability is assumed to be constant.  The results are qualitatively similar to the case of the steel vacuum vessel of the cryomodule, which has a similar thin walled cylindrical geometry, albeit on a larger scale.  The relative permeability of the steel pipe would be in the range 300 to 500 .

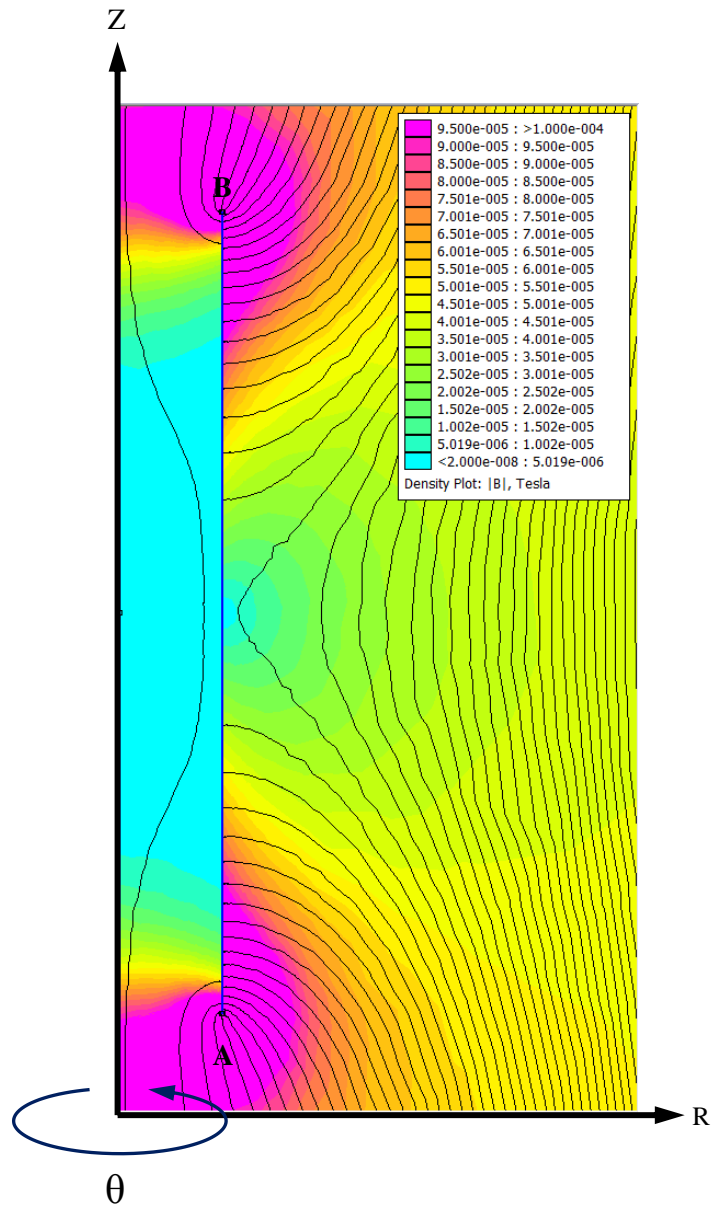

Figure 4   A High Permeability Cylinder Immersed In A Uniform Field Of 0.005 Tesla (500 milliGauss)

What is the field distribution within the walls of the cylinder?  The Z component of the field is plotted from cylinder point A to point B as the red curve in Figure 5.  The radial coordinate for the plot in Figure 5 is 127.5 mm, the midpoint of the cylinder walls.

After a *demagnetization* procedure, the cylinder will be left with a remanent internal field distribution that is similar to, but of smaller magnitude than, the magnetization due to $H_{DC}$.  The field in in the cylinder wall under the applied field is small and does not approach the saturated field limit for Cryoperm of ~ 0.7 Tesla, but there will be remanence, nonetheless [8].  Changes in field measurements of ± 5x10$^{-7}$ Tesla (5 milliGauss) have been observed, previous to this report, along the axis of Cryoperm cylindrical shields installed in a cryomodule vacuum vessel following *demagnetization* procedures at Fermilab.  Such changes, due to remanence, occur if the AC *demagnetization* cycle is terminated at too large a field value, for example, above ~0.65 Ampere-turns/meter.

A remanence model based on thin current sheets of equal magnitude, but opposite polarity, located on the inner and outer radii of the cylinder duplicates the field distribution due to $H_{DC}$ of Figure 4. The current in the sheets is then reduced until stray field from the cylinder measured along the axis is $5 \times 10^{-7}$ Tesla (5 milliGauss). This forms the basis for a model for remanence in the magnetic shield. The remanence in the cylinder wall that produces a 5 milliGauss central field is shown as a blue colored dashed line in Figure 5.

The axial field for this model is plotted in Figure 6. The associated field map due to remanence is shown in Figure 7. Arrows indicate field direction on a representative group of field lines. The reader should pay special attention to the field lines that return to the opposite pole within the interior of the cylinder. These will play an important role in the apparent ability of the shield to attenuate ambient field. The rectangle shown in blue colored lines around the cylinder is not physical. It is the boundary of a region of decreased mesh size for the finite element problem solver.

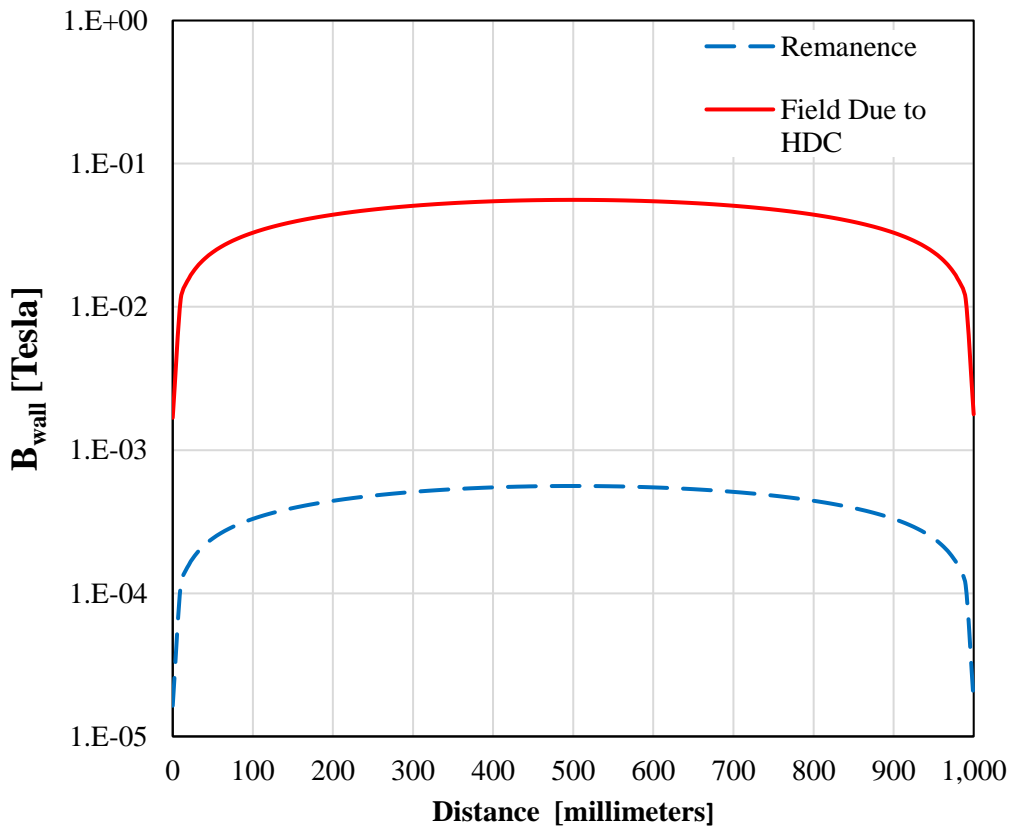

Figure 5    Field Distribution In The Cylinder Wall Due To Ambient Field = 0.005 Tesla And The Modeled Remanence

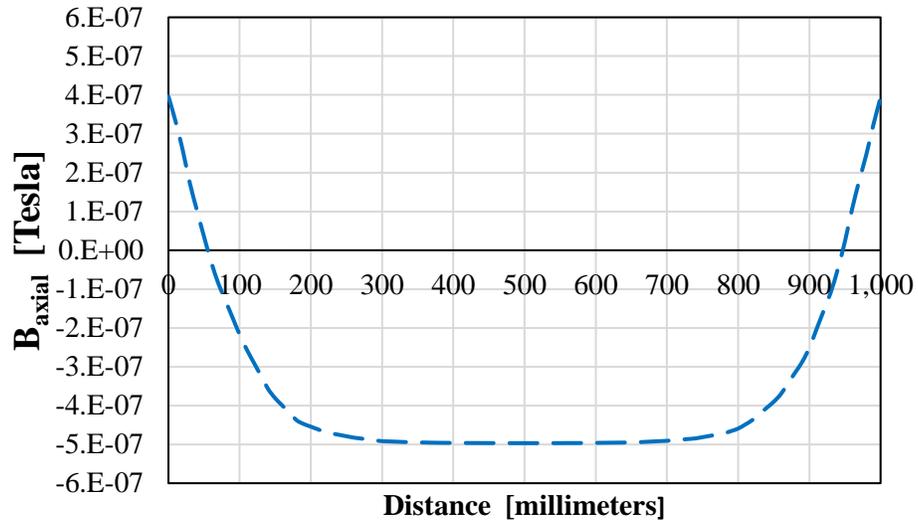

Figure 6    Field Along The Axis Of The Magnetized Cylinder Due To Remanence

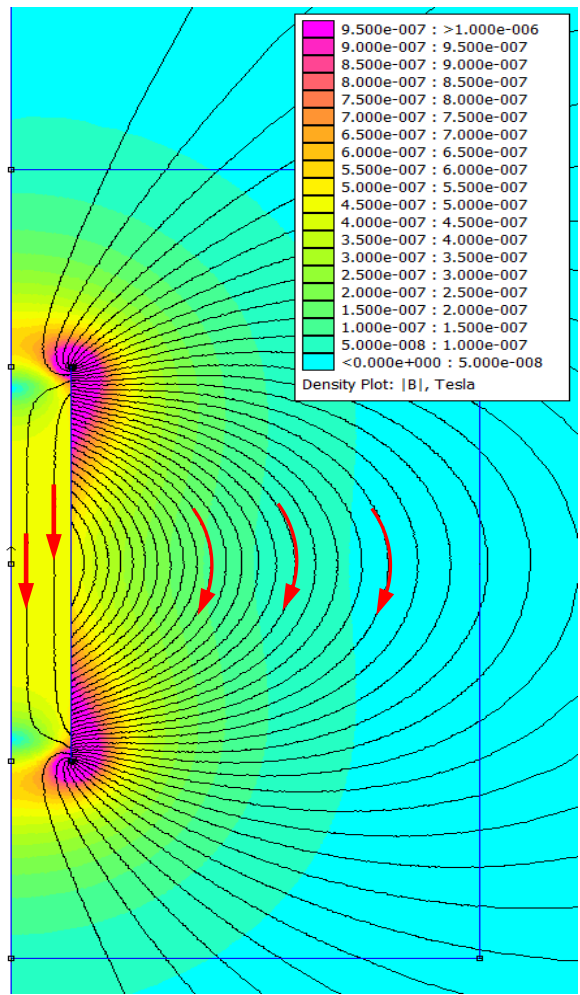

Figure 7    Magnetic Field Originating From The Magnetized Cylinder

## Shielding Efficiency

The model for internal magnetization after a *demagnetization* procedure now allows for investigation of the shielding efficiency of the cylinder. The following cases are considered:

Case 1. **Remanence Aligned**: The cylinder is immersed in the $5 \times 10^{-5}$ Tesla ambient field in which it was *demagnetized*

Case 2. **No Remanence**: A truly demagnetized cylinder, with zero remanence, in a $5 \times 10^{-5}$ Tesla ambient field

Case 3. **Remanence Anti-Aligned**: The cylinder is immersed in a $5 \times 10^{-5}$ Tesla ambient field of equal magnitude, but opposite sign from the field in which the cylinder was *demagnetized*

The field along the cylinder axis is shown in Figure 8 for each of the three cases. Distance is measured from the end of the open cylinder. The reader should note that there are large fields near the ends of the open cylinder. These fields diminish exponentially with distance along the cylinder axis toward the midpoint. It is the central portion of the length of the cylinder that is important to us here and that is indicative of the shielding efficiency of the material. A view of only the central portion of the cylinder is shown in the figure. In a real cryomodule end effects would be mitigated by the use of "end caps" on the cylinder, i.e., a local reduction in radius.

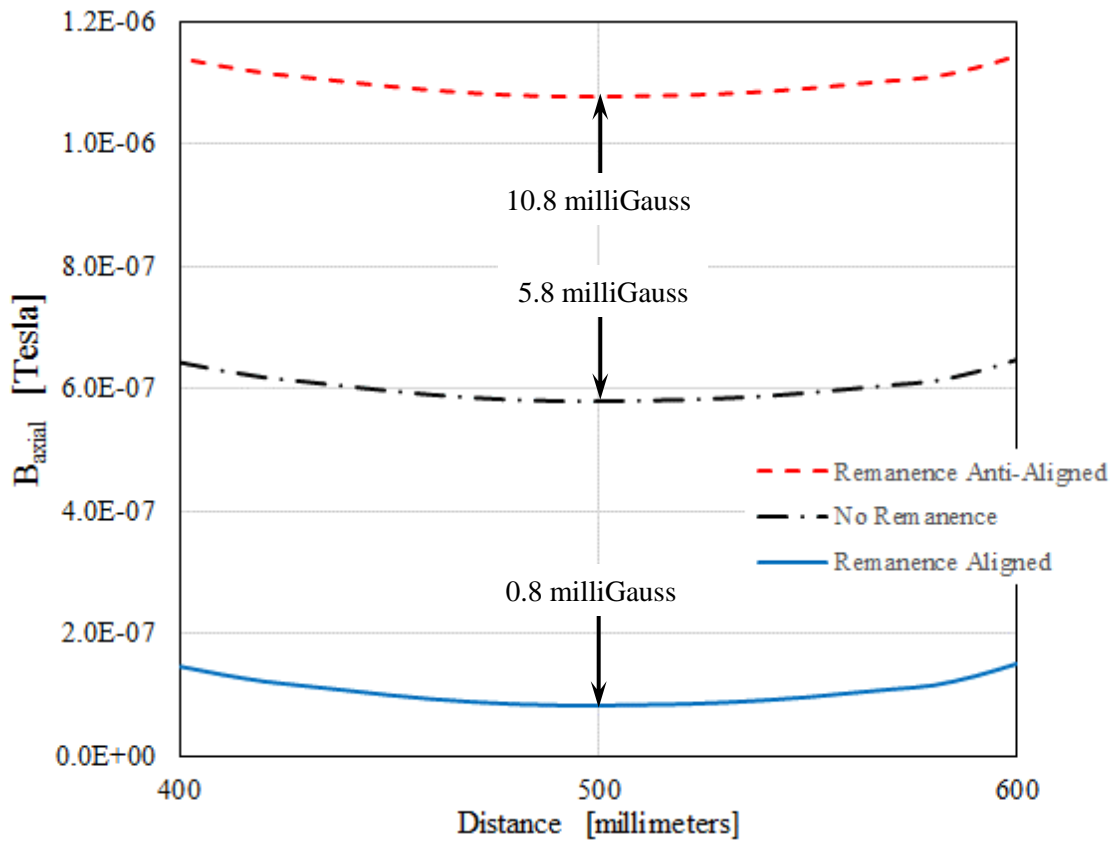

Figure 8  Attenuated Field For Different Orientations Of Remanence

In Figure 8 we see the influence of the internal stray field lines from Figure 7, which add to or subtract from, the attenuated ambient field.

## Discussion

The best result for field reduction at the SRF cavities is obtained by optimizing the remanence in shielding material and not from true demagnetizing. This is due to stray field lines originating from the remanence that return through the interior volume of the hollow cylinder, and would not be the case for a solid cylinder. In this context it is best to not think of hollow cylinders as simple dipoles. The modified flux return pattern matters a great deal for shielding applications.

The results shown here explain the experience of DESY/TTF researchers who in the 1990s made the observation that their test cryomodule "re-magnetized" itself when moved to another location. In reality, the magnetization of the cryomodule probably did not change. The ambient environmental field changed when the module was relocated, resulting in stray remanence field lines that were partially aligned with the new ambient field.

*Demagnetization* of the low carbon steel vacuum vessel is desirable for two different reasons. One reason is to optimize remanence as described in this note. Another important reason is to eliminate localized areas of large remanent field (values as high as $1 \times 10^{-3}$ Tesla) that are present after vendor operations of welding, bending, or handling of parts with electromagnets. High field (650 ampere-turns/meter) *demagnetization* is essential to achieve this latter goal. The optimization of remanence may be achieved with applied field lower than this value.

The next point addresses the question "Why not forget *demagnetization* and take care of the situation with field cancellation?" A cryomodule in a non uniform ambient field that has an average value of zero will not benefit from active field cancellation, unless there is a sufficiently large number of independent cancellation coils and associated field sensors (magnetometers) to allow smoothing the non uniformity. This works, but is complex and expensive. Even though the model in this report was made for a large object within a uniform ambient field, the description applies to local effects. This means that after *demagnetization* different points on a cryomodule within a non uniform ambient field will be locally optimized for attenuation, in effect, smoothing variations in the attenuated field. This, in turn, makes axial field cancellation more effective.

## Test Results

The cryomodule shown in Figure 9 was instrumented with an array of thirteen internal magnetometers. Unacceptably large changes in field, measured by these instruments, appeared after certain assembly procedures. At the time of this report welding procedures on the nearly complete cryomodule are being investigated as the source of the magnetization.

Table1 lists the readings of the magnetometers before and after a 350 Ampere-turns/meter remedial *demagnetization* procedure. Magnetometers numbered 1 through 8 were located in the space between the outer cavity wall and the inner wall of the helium vessel for four of the eight cavities in the cryomodule. Magnetometers 9 through 13 were located in the space between two layers of passive shielding around the outside of the helium vessel for five of the cavities. The magnetometers used were Bartington single axis fluxgates, model MAG-F, packaged for cryogenic applications. Their uncertainty is less than ± 0.01 µTesla.

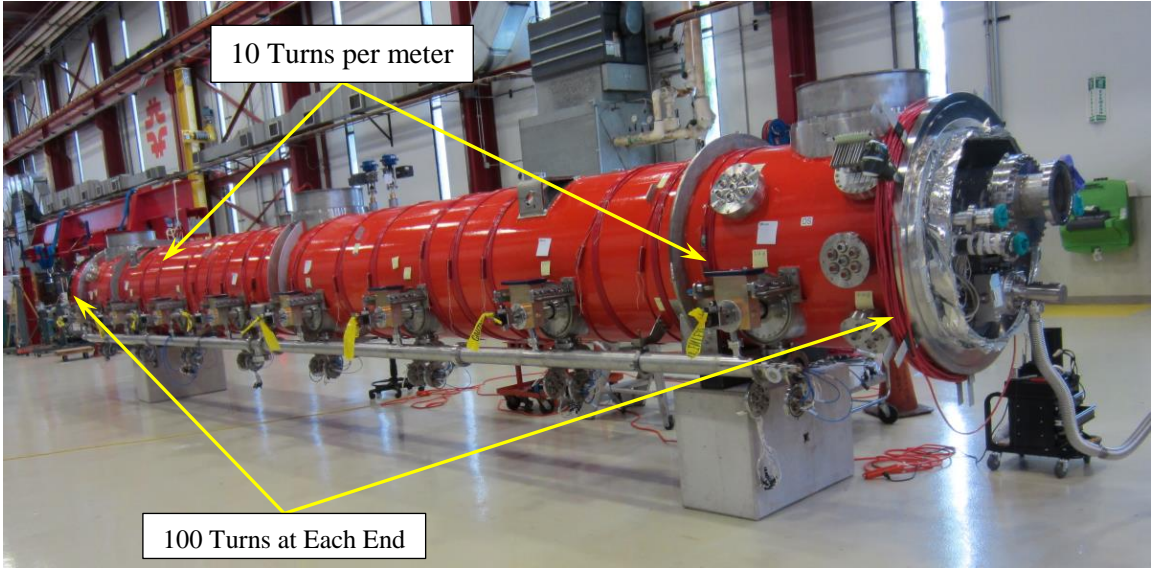

Figure 9   The cryomodule with *demagnetization* coils in place

| Magnetometer Number | Field before demagnetization [μTesla] | Field after demagnetization [μTesla] |
|---|---|---|
| 1 | -0.036 | -0.043 |
| 2 | -0.218 | 0.039 |
| 3 | -0.089 | -0.098 |
| 4 | -4.109 | -0.039 |
| 5 | -0.110 | 0.011 |
| 6 | 2.644 | 0.032 |
| 7 | 0.195 | -0.035 |
| 8 | -4.014 | -0.070 |
| 9 | -0.133 | 0.227 |
| 10 | -0.013 | 0.071 |
| 11 | 0.254 | 0.051 |
| 12 | 0.117 | 0.060 |
| 13 | -0.358 | 0.099 |

Table 1   Effect of *demagnetization* on field measurements

From Table 1 we see that field with a magnitude as high as $4.1 \times 10^{-6}$ Tesla (40 milliGauss) is reduced to the level of $0.1 \times 10^{-6}$ Tesla (1 milliGauss) by the procedure. Indeed, there are no readings greater than $0.23 \times 10^{-6}$ Tesla (2.3 milliGauss) following the *demagnetization*. It should be re-emphasized that the improvement is due to two different phenomena that proceed simultaneously during cryomodule *demagnetization*: (1) the reduction of local areas of large remanent field in the cryomodule and (2) the adjustment of the cryomodule remanence to the local ambient environmental field. Therefore, *demagnetization* is seen to be an essential step in achieving satisfactorily low field for nitrogen doped cavity cryomodules until the time when the technology of flux expulsion has been perfected.

## Conclusion

Stray field from remanent magnetization in a thin walled ferromagnetic cylindrical shield adds vectorially to the attenuated ambient field at the location of the object that is to be shielded. In situ *demagnetization* optimizes the direction of the remanent field for shielding purposes. If the magnetic environment of a cryomodule is changed, it will be beneficial to repeat the *demagnetization* procedure for the new environment. These effects have now been successfully demonstrated in a fully assembled 11.4 meter cryomodule.

## Acknowledgement

An effective and affordable polarity reversing power supply was created to match the requirements of this project by M. Matulik and M. Cherry of the Fermilab Particle Physics Division. A description of this device will appear in a future publication. A fully detailed description of the *demagnetization* of the cryomodule shown in Figure 9 is forthcoming.

___________________________________________________________________________________

## References

[1] A. Grassellino, A. Romanenko, O. Melnychuk, Y. Trenikhina, A. Crawford, A. Rowe, M. Wong, D. Sergatskov, T. Khabiboulline, and F. Barkov, " Nitrogen and argon doping of niobium for superconducting radio frequency cavities: A pathway to highly efficient accelerating structures," Supercond. Sci. Technol. 26, 102001 (2013). http://dx.doi.org/10.1088/0953-2048/26/10/102001

[2] H. Padamsee, J. Knobloch, and T. Hays, RF Superconductivity for Accelerators ( Wiley, New York, 1998), p. 173

[3] "The Conceptual Design Report for the TeSLA Test Facility Linac: Version 1.0", 1995, p. 160 http://tesla.desy.de/TTF_Report/CDR/pdf/cdr_chap4.pdf

[4] A. Crawford, "In situ cryomodule demagnetization", https://arxiv.org/ftp/arxiv/papers/1507/1507.06582.pdf

[5] A. Crawford, "A study of thermocurrent induced magnetic field in ILC cavities", https://arxiv.org/ftp/arxiv/papers/1403/1403.7996.pdf

[6] M. Martinello, M. Cecchin, A. Grassellino, A. Crawford, O. Melnychuk, A. Romanenko, D. Sergatskov, "Magnetic flux studies in horizontally cooled elliptical superconducting cavities", J. Appl. Phys. 118, 044505 (2015); http://dx.doi.org/10.1063/1.4927519

[7] S. Posen, M. Cecchin, A. Crawford, A. Grassellino, M. Martinello, O. Melnychuk, A. Romanenko, D. Sergatskov, Y. Trenikhina, "Efficient expulsion of magnetic flux in superconducting radiofrequency cavities for high Q0 applications", J. Appl. Phys. 119, 213903 (2016); http://dx.doi.org/10.1063/1.4953087

[8 ] Amuneal Corp. Promotional Literature: http://www.amuneal.com/wp-content/uploads/2016/05/AmunealDataSheet2.pdf